# Current-induced motion of a transverse magnetic domain wall in the presence of spin Hall effect

Soo-Man Seo[1], Kyoung-Whan Kim[2], Jisu Ryu[2], Hyun-Woo Lee[2,a)], and Kyung-Jin Lee[1,3,4,b)]

[1]*Department of Materials Science and Engineering, Korea University, Seoul 136-701, Korea*

[2]*PCTP and Department of Physics, Pohang University of Science and Technology, Kyungbuk 790-784, Korea*

[3]*Center for Nanoscale Science and Technology, National Institute of Standards and Technology, Gaithersburg, Maryland 20899-8412, USA*

[4]*Maryland Nanocenter, University of Maryland, College Park, MD 20742, USA*

**We theoretically study the current-induced dynamics of a transverse magnetic domain wall in bi-layer nanowires consisting of a ferromagnet on top of a nonmagnet having strong spin-orbit coupling. Domain wall dynamics is characterized by two threshold current densities, $J_{th}^{WB}$ and $J_{th}^{REV}$, where $J_{th}^{WB}$ is a threshold for the chirality switching of the domain wall and $J_{th}^{REV}$ is another threshold for the reversed domain wall motion caused by spin Hall effect. Domain walls with a certain chirality may move opposite to the electron-flow direction with high speed in the current range $J_{th}^{REV} < J < J_{th}^{WB}$ for the system designed to satisfy the conditions $J_{th}^{WB} > J_{th}^{REV}$ and $\alpha > \beta$, where $\alpha$ is the Gilbert damping constant and $\beta$ is the nonadiabaticity of spin torque. Micromagnetic simulations confirm the validity of analytical results.**

[a)]Electronic mail: hwl@postech.ac.kr.
[b)]Electronic mail: kj_lee@korea.ac.kr.

Electric manipulation of domain walls (DWs) in magnetic nanowires can be realized by the spin-transfer torque (STT) due to the coupling between local magnetic moments of the DW and spin-polarized currents[1,2]. Numerous studies on this subject have addressed its fundamental physics[3-5], and to explore its potential in applications such as data storage and logic devices[6]. Up until now, however, most studies have focused on the effect of the spin current that is polarized by a ferromagnetic layer.

Another way to generate a spin current is the spin Hall effect (SHE)[7,8]. In ferromagnet (FM)|nonmagnet (NM) bi-layer systems, an in-plane charge current density ($J_c$) passing through the NM is converted into a perpendicular spin current density ($J_s$) owing to the SHE. The ratio of $J_s$ to $J_c$ is parameterized by spin Hall angle. This spin current caused by SHE exerts a STT (= SHE-STT) on the FM and consequently modifies its magnetization dynamics. During the last decade, most studies on the SHE have focused on measuring the spin Hall angle[9-14]. Recently the magnetization switching[15] and the modulation of propagating spin waves by SHE-STT were investigated[16-18]. However, the effect of SHE-STT on current-induced DW dynamics has not been treated.

In this Letter, we study DW dynamics including all current-induced STTs in a nanowire consisting of FM/NM bi-layers (Fig. 1), where FM has an in-plane magnetic anisotropy and NM has strong spin-orbit coupling (SOC) responsible for the SHE. A charge current passing through the FM generates conventional adiabatic and nonadiabatic STTs[19-21], whereas a charge current flowing through the NM experiences SHE and generates SHE-STT on the FM. For the current running in the $x$ axis, the modified Landau-Lifshitz-Gilbert equation including all the STTs is given by

$$\frac{\partial \mathbf{m}}{\partial t} = -\gamma \mathbf{m} \times \mathbf{H}_{\text{eff}} + \alpha \mathbf{m} \times \frac{\partial \mathbf{m}}{\partial t} - b_J \mathbf{m} \times \left( \mathbf{m} \times \frac{\partial \mathbf{m}}{\partial x} \right) - \beta b_J \mathbf{m} \times \frac{\partial \mathbf{m}}{\partial x} - \theta_{SH} c_J \mathbf{m} \times (\mathbf{m} \times \hat{y}), \quad (1)$$

where **m** is the unit vector along the magnetization, $\alpha$ is the Gilbert damping constant, $b_J$ $(= g\mu_B P J_F/2eM_S)$ is the magnitude of adiabatic STT, $\beta$ is the nonadiabaticity of STT, $\theta_{SH} c_J$ $(= \theta_{SH} \gamma \hbar J_N/2eM_S t_F)$ is the magnitude of SHE-STT, $\theta_{SH}$ is an effective spin Hall angle for the bi-layer system, $\gamma$ is the gyromagnetic ratio, $g$ is the Landé $g$-factor, $\mu_B$ is the Bohr magneton, $P$ is the spin polarization in the FM, $e$ is the electron charge, $M_S$ is the saturation magnetization of the FM, and $J_F$ ($J_N$) is the current density in the FM (NM). $J_F$ and $J_N$ are determined by a simple circuit model; i.e., $J_F = J_0(t_F + t_N)\sigma_F/(t_F \sigma_F + t_N \sigma_N)$ and $J_N = J_0(t_F + t_N)\sigma_N/(t_F \sigma_F + t_N \sigma_N)$, where $J_0$ is the total current density in the bi-layer nanowire, $\sigma_F$ ($\sigma_N$) is the conductivity of the FM (NM), and $t_F$ ($t_N$) is the thickness of the FM (NM). We assume that $\theta_{SH}$ is smaller than 1 as is usually the case experimentally.

For a nanowire with an in-plane magnetic anisotropy, a net effective field is given by

$$\mathbf{H}_{eff} = \frac{2A}{M_S}\frac{\partial^2 \mathbf{m}}{\partial x^2} + H_K m_x \hat{\mathbf{x}} + \mathbf{H}_d, \qquad (2)$$

where $A$ is the exchange stiffness constant, $H_K$ is the easy axis anisotropy field along the $x$ axis, and $\mathbf{H}_d$ is the magnetostatic field given by $\mathbf{H}_d(\mathbf{r}) = M_S \int d^3 \mathbf{r}' \tilde{N}(\mathbf{r} - \mathbf{r}')\mathbf{m}(\mathbf{r}')$, where the components of the tensor $\tilde{N}$ are given by $N_{xx} = -[1 - 3x^2/|\mathbf{r}|^2]/|\mathbf{r}|^3$, $N_{xy} = 3xy/|\mathbf{r}|^5$ [22]. Other components are defined in a similar way. For a one-dimensional DW as shown in Fig. 1, the spatial profile of the magnetization is described by $\mathbf{m} = (\cos\theta, \sin\theta\cos\phi, \sin\theta\sin\phi)$, where $\sin\theta = \text{sech}[(x-X)/\lambda]$, $\cos\theta = \tanh[(x-X)/\lambda]$, $X(t)$ is the DW position, $\phi(t)$ is the DW tilt angle, and $\lambda$ is the DW width. By using the procedure developed by Thiele[23], we obtain the equations of motion for the two collective coordinates $X$ and $\phi$ in the rigid DW limit,

$$-\frac{\partial X}{\partial t} + \alpha\lambda\frac{\partial \phi}{\partial t} = b_J - \frac{\gamma H_d \lambda}{2}\sin(2\phi), \qquad (3)$$

$$\lambda \frac{\partial \phi}{\partial t} + \alpha \frac{\partial X}{\partial t} = -\beta_{eff} b_J, \tag{4}$$

where $H_d(=2K_d/M_S)$, $\beta_{eff} = \beta(1 + B_{SH}\lambda \sin\phi)$, $B_{SH} = \pi\theta_{SH} J_N / 2\beta t_F P J_F$, and $K_d$ is the hard-axis anisotropy energy density. From Eqs. (3) and (4), one finds that the effect of SHE-STT on DW dynamics is captured by replacing $\beta$ by $\beta_{eff}$. Assuming that FM is Permalloy (Py: $Ni_{80}Fe_{20}$) and NM is Pt, for the parameters of $t_F$ = 4 nm, $t_N$ = 3 nm, $\sigma_F = \sigma_N$, $\theta_{SH}$ = 0.1, $\beta \approx$ 0.01 to 0.03 [24], $P$ = 0.7, and $\lambda$ = 30 nm, we find $B_{SH}\lambda \approx$ 18 to 56, which is not small. Therefore, $\beta_{eff}$ can be much larger than $\beta$ unless $\sin\phi$ is extremely small. Furthermore, it is possible that $\beta_{eff}$ is even negative if $B_{SH}\lambda \sin\phi < -1$.

To get an insight into the effect of SHE-STT on DW dynamics, we derive several analytical solutions from Eqs. (3) and (4). It is known that DW dynamics in a nanowire can be classified into two regimes; i.e., below and above the Walker breakdown[25]. Below the Walker breakdown, $\phi$ increases in the initial time stage and then becomes saturated to a certain value over time. In this limit ($\partial\phi/\partial t = 0$ as $t \to \infty$), we obtain

$$b_J = \frac{\gamma \alpha H_d \lambda \sin 2\phi}{2(\alpha - \beta_{eff})}, \tag{5}$$

Threshold adiabatic STT for the Walker breakdown ($b_J^{WB}$) is obtained from the maximum value of the right-hand-side of Eq. (5); i.e., $b_J^{WB} = \max[\gamma\alpha H_d \lambda \sin 2\phi / 2(\alpha - \beta_{eff})]$. Note that $b_J^{WB}$ is not simply $[\gamma\alpha H_d \lambda / 2(\alpha - \beta_{eff})]$ because $\beta_{eff}$ also includes $\phi$. When $B_{SH} = 0$, Eq. (5) reduces to $b_J^{WB} = \gamma\alpha H_d \lambda / 2(\alpha - \beta)$, reproducing the previous result [26] in the absence of SHE.

For $|b_J| < |b_J^{WB}|$ (below the Walker breakdown) and using the small-angle approximation, DW velocity ($v_{DW}$) is given by

$$v_{DW} = -\frac{\beta}{\alpha} b_J \left( 1 \pm B_{SH} \frac{(\alpha - \beta) b_J}{\gamma \alpha H_d \pm \beta b_J B_{SH}} \right), \quad (6)$$

where the sign "+" and "−" in the parenthesis corresponds to the initial tilt angles $\phi_0 = \pi$ and $\phi_0 = 0$, respectively. This $\phi_0$ dependence of $v_{DW}$ originates from the fact that SHE-STT acts like a damping or an anti-damping term depending on $\phi_0$. When $\beta = \alpha$, $v_{DW} = -b_J$ so that $v_{DW}$ does not depend on SHE-STT. However, this condition is hardly realized in the bi-layer system that we consider since the strong SOC in NM increases the intrinsic $\alpha$ of FM through the spin pumping effect[27]. When $B_{SH} = 0$, $v_{DW} = -(\beta/\alpha)b_J$, consistent with the DW velocity in the absence of SHE[26]. Note that in our sign convention, a negative $b_J$ corresponds to the electron-flow in +x direction and a positive $v_{DW}$ corresponds to the DW motion along the electron-flow direction. Therefore, when the term in the parenthesis of Eq. (6) is negative, the DW moves against the electron-flow direction instead of along it. Threshold adiabatic STT for this reversed DW motion ($b_J^{REV}$) is given by

$$b_J^{REV} = \mp \frac{\gamma H_d}{B_{SH}}. \quad (7)$$

For $|b_J| \gg |b_J^{WB}|$ (far above the Walker breakdown), the time-averaged values of $\sin\phi$ and $\sin 2\phi$ can be set to zero because of the precession of $\phi$. In this limit, $v_{DW}$ is determined by Eq. (3) and becomes $-b_J$ so that the DW moves along the electron-flow direction and its motion does not depend on SHE-STT.

Based on the above investigations, there are two interesting effects of SHE on current-induced DW dynamics. First, current-induced DW dynamics is determined by two thresholds,

$b_J^{WB}$ and $b_J^{REV}$. When $|b_J^{REV}| < |b_J| < |b_J^{WB}|$, the DW can move against the electron-flow direction. Note that the existence of such $b_J$ range implicitly assumes $|b_J^{REV}| < |b_J^{WB}|$. When this inequality is not satisfied, the DW always moves along the electron-flow direction. For all cases, $|v_{DW}|$ can be larger than $|\beta b_J/\alpha|$ depending on the parameters (see Eq. (6)). Second, $v_{DW}$ is asymmetric against the initial tilt angle $\phi_0$ for a fixed current polarity. A similar argument is also valid for a fixed $\phi_0$ but with varying the current polarity; i.e., $v_{DW}$ is asymmetric with respect to the current polarity for a fixed $\phi_0$. This behavior follows because SHE-STT acts like a damping term for one sign of the current but acts like an anti-damping term for the other sign. Therefore, although the condition of $|b_J^{REV}| < |b_J| < |b_J^{WB}|$ is satisfied, the reversed DW motion is expected to be observed only for one current polarity.

To verify the analytical results, we perform a one-dimensional micromagnetic simulation by numerically solving Eq. (1). We consider a Py/Pt bi-layer nanowire of (length × width × thickness) = (2000 nm × 80 nm × 4 nm (Py) and 3 nm (Pt)) (Fig. 1). Py material parameters of $M_S$ = 800 kA/m, $A$ = 1.3×10$^{-11}$ J/m, $P$ = 0.7, $\alpha$ = 0.02, and $\beta$ = 0.01 to 0.03 are used. The crystalline anisotropy and the temperature are assumed to be zero. Conductivities of both layers are assumed to be the same as $\sigma_{Py}$ = $\sigma_{Pt}$ = 6.5 (μΩm)$^{-1}$, and thus $J_0 = J_F = J_N$. For all cases, the initial DW tilt angle $\phi_0$ is set to zero.

Analytical and numerical results are compared in Fig. 2. DW velocity ($v_{DW}$) and DW tilt angle ($\phi_{DW}$) as a function of the total current density of the bi-layer ($J_0$) for three values of $\theta_{SH}$ (= +0.1, 0.0, −0.1) and $\beta$ = 0.01 (thus $\alpha > \beta$) are shown in Fig. 2(a) and (b), respectively. $v_{DW}$ is estimated from the terminal velocity. Here, we test both positive and negative values of $\theta_{SH}$ since the spin Hall angle can have either sign.

Current dependences of $v_{DW}$ (Fig. 2(a)) and $\phi_{DW}$ (Fig. 2(b)) show close correlation, meaning that the DW tilting plays a crucial role for the effect of SHE on DW dynamics as demonstrated analytically. In Fig. 2(a), the numerical results (symbols) are in agreement with the results obtained from Eq. (6) (lines). For $\theta_{SH} = 0$, $v_{DW}$ is linearly proportional to $J_0$ and the DW always moves along the electron-flow direction. However, for $0.5 \times 10^{12} \leq J_0 \leq 1.0 \times 10^{12}$ A/m² with $\theta_{SH} = -0.1$ ($-1.0 \times 10^{12}$ A/m² $\leq J_0 \leq -0.5 \times 10^{12}$ A/m² with $\theta_{SH} = 0.1$), $v_{DW}$ has the same polarity as the current. Thus, the DW moves along the current-flow direction for these ranges of the current. The threshold for the reversed DW motion is consistent with the analytical solution of Eq. (7); i.e., $b_J^{REV} = \pm 26.6$ m/s corresponding to $J_0 = \pm 0.52 \times 10^{12}$ A/m². The maximum $v_{DW}$ is obtained at $J_0 = \pm 1.0 \times 10^{12}$ A/m² immediately before the DW experiences Walker breakdown and switches its chirality. As shown in the Fig. 2 (c), the normalized y-component of the magnetization at the DW center ($m_y$) abruptly changes from $\pm 1$ to $\mp 1$ for $J_0 = \pm 1.0 \times 10^{12}$ A/m² and $\theta_{SH} = \mp 0.1$. This current density is consistent with the threshold for Walker breakdown ($b_J^{WB}$); i.e., $b_J^{WB} = \pm 53$ m/s corresponding to $J_0 = \pm 1.045 \times 10^{12}$ A/m². At this current density, $v_{DW}$ is enhanced by a factor of 5 compared to the case for $\theta_{SH} = 0$.

Fig. 3 (a) and (b) show $v_{DW}$ and $\phi_{DW}$ as a function of $J_0$ for three values of $\theta_{SH}$ (= +0.1, 0.0, −0.1) and $\beta = 0.03$ (thus $\alpha < \beta$). Similarly to the cases for $\beta = 0.01$, $v_{DW}$ is closely correlated to $\phi_{DW}$ and significantly enhanced near $b_J^{WB}$. In this case, in contrast to the case for $\alpha > \beta$, reversed DW motion is not observed. It is because the sign of the $(\alpha - \beta)$ term in Eq. (6) is negative in this case, and thus the overall sign of $v_{DW}$ corresponds to the DW motion along the electron-flow direction. We find that the current-induced Oersted field has only a negligible effect on $v_{DW}$ (not shown). Thus, the numerical results confirm the validity of the analytical solutions; the DW moves along the current-flow direction at the limited range of the current (i.e., $\left|b_J^{REV}\right|<\left|b_J\right|<\left|b_J^{WB}\right|$) when $\alpha>\beta$. In addition this reversed DW motion appears only for one current polarity.

Finally, we remark the effect of SHE on DW dynamics in the nanowire with a perpendicular anisotropy. It was experimentally reported that the DW moves along the current-flow direction with a high $v_{DW}$ ($\approx$ 400 m/s) in the perpendicularly magnetized nanowire consisting of Pt/Co/AlO$_x$ [28, 29]. We note that this DW dynamics cannot be explained by the SHE only. Considering the materials parameters in Ref. [29] as $M_S$ = 1090 kA/m, $K$ = 1.2×10$^6$ J/m$^3$, $A$ = 1.3×10$^{-11}$ J/m, $\alpha$ = 0.2, $P$ = 0.7, $\lambda$ = 5 nm, and assuming $\theta_{SH}$ = 0.1 and $\beta$ = 0.1, we find $B_{SH}\lambda$ = 18.8 that is comparable to the value for the Py/Pt bi-layer tested in this work. For Pt/Co/AlO$_x$, however, $b_J^{REV}$ and $b_J^{WB}$ are respectively −1.5 and −3 m/s (corresponding to $J_0$ = −0.4×10$^{11}$ and −0.8×10$^{11}$ A/m$^2$). These thresholds are much smaller than those of the Py/Pt bi-layer since $H_d$ of DW in a perpendicular system is smaller than in an in-plane system (i.e., $H_d$ = 848 mT for the system of Py/Pt in this work, 33 mT for the system in Ref. [29])[22]. Note that the maximum DW velocity moving along the current-flow direction ($v_{DW}^{REV}$) is obtained at $b_J=b_J^{WB}$. The $b_J^{WB}$ (= −3 m/s) in Pt/Co/AlO$_x$ system is too small to allow such a high $v_{DW}^{REV}$ ($\approx$ −400 m/s). Indeed, the numerically obtained maximum $v_{DW}^{REV}$ is −8.2 m/s at $J_0$ = −0.71×10$^{11}$ A/m$^2$ ($b_J$ = −2.64 m/s) (not shown), which is much smaller than the experimentally obtained value, −400 m/s. More importantly, in the Pt/Co/AlO$_x$ system, the reversed DW motion was observed at both current polarities[31] whereas the SHE allows the reversed motion at only one current polarity. On the other hand, we theoretically demonstrated that the DW dynamics reported in Ref. [28, 29] can be explained by STTs caused by Rashba SOC[32]. We also remark that one of us reported the effect of SOC on current-driven DW motion recently[33]. In Ref. [33], however, the effect of SOC within FM was investigated, in contrast to the present work where the effect of SOC in NM of the FM/NM bi-layer system is investigated.

To conclude, we present the analytical model for current-induced DW motion in the presence of SHE. We demonstrate that DW dynamics is significantly affected by the SHE. In particular, for the case of $\alpha > \beta$, the SHE enables the reversed DW motion with high speed at one current polarity when the system is designed to satisfy the condition of $\left|b_J^{REV}\right| < \left|b_J^{WB}\right|$ and the current density is selected to be in the range between the two thresholds. Our result demonstrates that the engineering of SOC and thus the SHE provides an important opportunity for an efficient operation of spintronic devices.

This work was supported by the NRF (2010-0014109, 2010-0023798, 2011-0009278, 2011-0028163, 2011-0030046) and the MKE/KEIT (2009-F-004-01). K.J.L. acknowledges support under the Cooperative Research Agreement between the University of Maryland and the National Institute of Standards and Technology Center for Nanoscale Science and Technology, Award 70NANB10H193, through the University of Maryland.

**REFERENCE**

[1] J. C. Slonczewski, J. Magn. Mag. Mater. **159**, L1 (1996).

[2] L. Berger, Phys. Rev. B **54**, 9353 (1996).

[3] A. Yamaguchi, T. Ono, S. Nasu, K. Miyake, K. Mibu, and T. Shinjo, Phys. Rev. Lett. **92**, 077205 (2004).

[4] M. Yamanouchi, D. Chiba, F. Matsukura, and H. Ohno, Nature (London) **428**, 539 (2004).

[5] M. Kläui, C. A. F. Vaz, J. A. C. Bland, W. Wernsdorfer, G. Faini, E. Cambril, L. J. Heyderman, F. Nolting, and U. Rüdiger, Phys. Rev. Lett. **94**, 106601 (2005).

[6] S. S. P. Parkin, M. Hayashi, and L. Thomas, Science **320**, 190 (2008).

[7] J. E. Hirsch, Phys. Rev. Lett. **83**, 1834 (1999).

[8] S. Zhang, Phys. Rev. Lett. **85**, 393 (2000).

[9] S. O. Valenzuela and M. Tinkham, Nature (London) **442**, 176 (2006).

[10] T. Kimura, Y. Otani, T. Sato, S. Takahashi, and S. Maekawa, Phys. Rev. Lett. **98**, 156601 (2007).

[11] K. Ando, S. Takahashi, K. Harii, K. Sasage, J. Ieda, S. Maekawa, and E. Saitoh, Phys. Rev. Lett. **101**, 036601 (2008).

[12] T. Seki, Y. Hasegawa, S. Mitani, S. Takahashi, H. Imamura, S. Maekawa, J. Nitta, and K. Takahashi, Nat. Mater. **7**, 125 (2008).

[13] O. Mosendz, J. E. Pearson, F. Y. Fradin, G. E. W. Bauer, S. D. Bader, and A. Hoffman, Phys. Rev. Lett. **104**, 046601 (2010).

[14] L. Liu, T. Moriyama, D. C. Ralph, and R. A. Buhrman, Phys. Rev. Lett. **106**, 106602 (2011).

[15] L. Liu, O. J. Lee, T. J. Gudmundsen, D. C. Ralph, and R. A. Buhrman, arXiv:1110.6846.

[16] V. E. Demidov, S. Urazhdin, E. R. J. Edwards, M. D. Stiles, R. D. McMichael, and S. O.

Demokritov, Phys. Rev. Lett. **107**, 107204 (2011).

[17] Z. Wang, Y. Sun, M. Wu, V. Tiberkevich, and A. Slavin, Phys. Rev. Lett. **107**, 146602 (2011).

[18] E. Padrón-Hernández, A. Azevedo, and S. M. Rezende, Appl. Phys. Lett. **99**, 192511 (2011).

[19] G. Tatara and H. Kohno, Phys. Rev. Lett. **92**, 086601 (2004).

[20] S. Zhang and Z. Li, Phys. Rev. Lett. **93**, 127204 (2004).

[21] A. Thiaville, Y. Nakatani, J. Miltat, and Y. Suzuki, Europhys. Lett. **6**9, 990 (2005).

[22] S.-W. Jung, W. Kim, T.-D. Lee, K.-J. Lee, and H.-W. Lee, Appl. Phys. Lett. **92**, 202508 (2008).

[23] A. A. Thiele, Phys. Rev. Lett. **30**, 230 (1973).

[24] K. Sekiguchi, K. Yamda, S.-M. Seo, K.-J. Lee, D. Chiba, K. Kobayashi, and T. Ono, Phys. Rev. Lett. **108**, 017203 (2012).

[25] N. L. Schryer and L. R. Walker, J. Appl. Phys. **45**, 5406 (1974).

[26] A. Mougin, M. Cormier, J. P. Adam, P. J. Metaxas, and J. Ferré, Europhys. Lett. **78**, 57007 (2007).

[27] Y. Tserkovnyak and A, Brataas, and G. E. Bauer, Phys. Rev. Lett. **88**, 117601 (2002).

[28] T. A. Moore, I. M. Miron, G. Gaudin, G. Serret, S. Auffret, B. Rodmacq, A. Schul, S. Pizzini, J. Vogel, and M. Bonfim, Appl. Phys. Lett. **93**, 262504 (2008); *ibid* **95**, 179902 (2009).

[29] I. M. Miron, T. Moore, H. Szambolics, L. D. Buda-Prejbeanu, S. Auffret, B. Rodmacq, S. Pizzini, J. Vogel, M. Bonfim, A. Schul, and G. Gaudin, Nat. Mater. **10**, 189 (2011).

[31] I. M. Miron, private communication.

[32] K.-W. Kim, S.-M. Seo, J. Ryu, K.-J. Lee, and H.-W. Lee, arXiv:1111.3422v2.

[33] A. Manchon and K.-J. Lee, Appl. Phys. Lett. **99**, 022504 (2011); *ibid* **99**, 229905 (2011).

**FIGURE CAPTION**

FIG. 1. (Color online) Schematics of FM/NM bi-layer nanowire. (top) Structure. (lower left) Spatial profile of DW. The colored contour shows $x$ component of the magnetization for 2-D micromagnetics. (lower right) Width-averaged magnetization components.

FIG. 2. (Color online) Domain wall velocity for $\alpha > \beta$. (a) DW velocity ($v_{DW}$) as a function of the total current density of bi-layer ($J_0$) for three values of $\theta_{SH}$ (= +0.1, 0.0, −0.1) and $\beta = 0.01$ ($\alpha = 0.02$). Symbols are modeling results, whereas solid lines correspond to Eq. (6). (b) DW tilt angle ($\phi_{DW}$) as a function of $J_0$. Filled symbols above the chiral switching threshold ($|J_0| = 1.1 \times 10^{12}$ A/m$^2$) are shifted from their original values by −180° (filled green triangles) and +180° (filled red circles). (c) Normalized $y$ component of the magnetization at the DW center ($m_y$) as a function of $J_0$.

FIG. 3. (Color online) Domain wall velocity for $\alpha < \beta$. (a) DW velocity ($v_{DW}$) as a function of the total current density of bi-layer ($J_0$) for three values of $\theta_{SH}$ (= +0.1, 0.0, −0.1) and $\beta = 0.03$ ($\alpha = 0.02$). Symbols are modeling results, whereas solid lines correspond to Eq. (6). (b) DW tilt angle ($\phi_{DW}$) as a function of $J_0$. Filled symbols represent the cases that the chirality of DW switches from its initial tilt angle $\phi_0 = 0$. (c) Normalized $y$ component of the magnetization at the DW center ($m_y$) as a function of $J_0$.

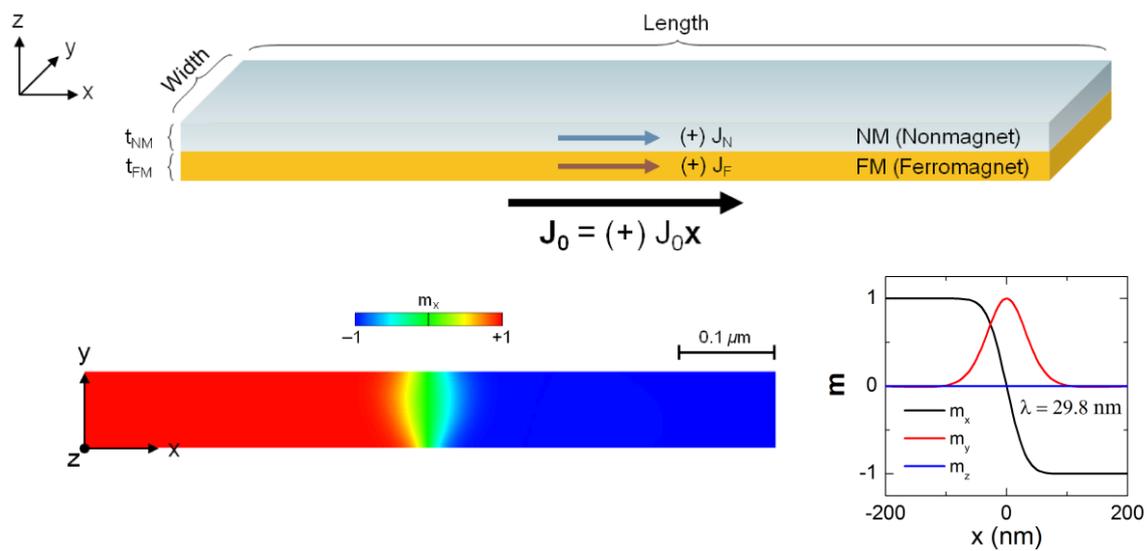

FIG. 1. Seo *et al.*

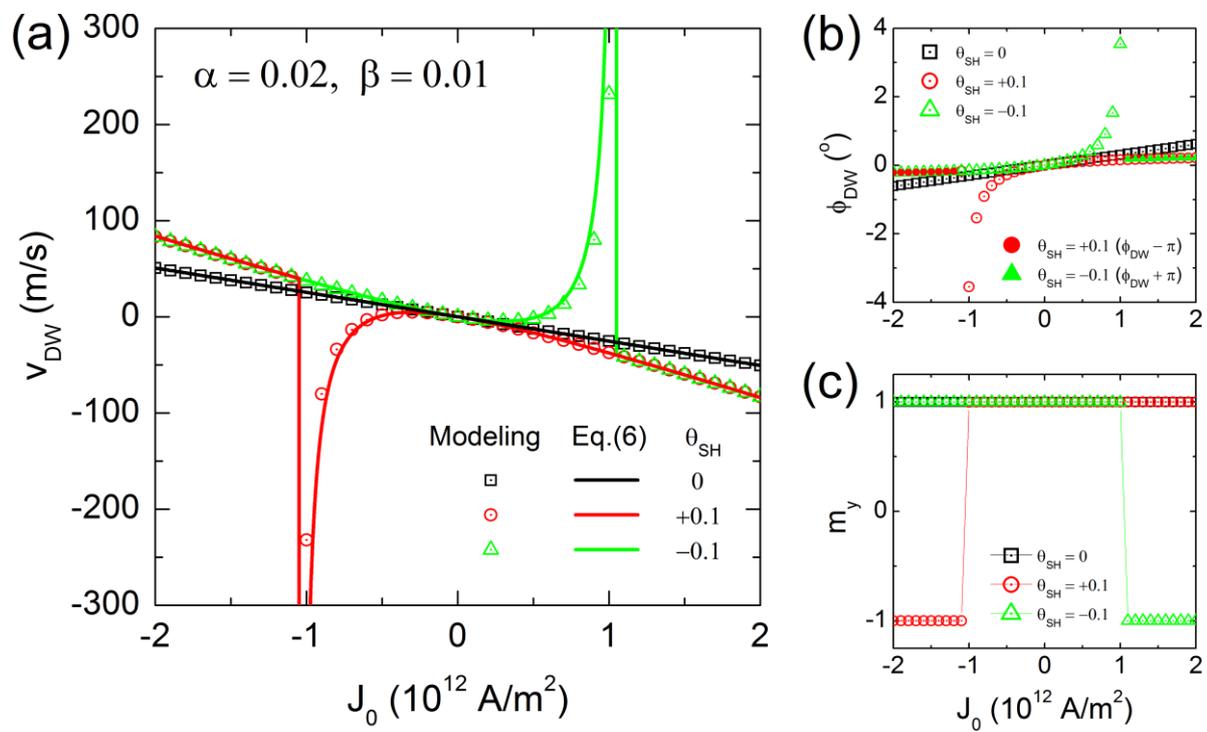

FIG. 2. Seo *et al.*

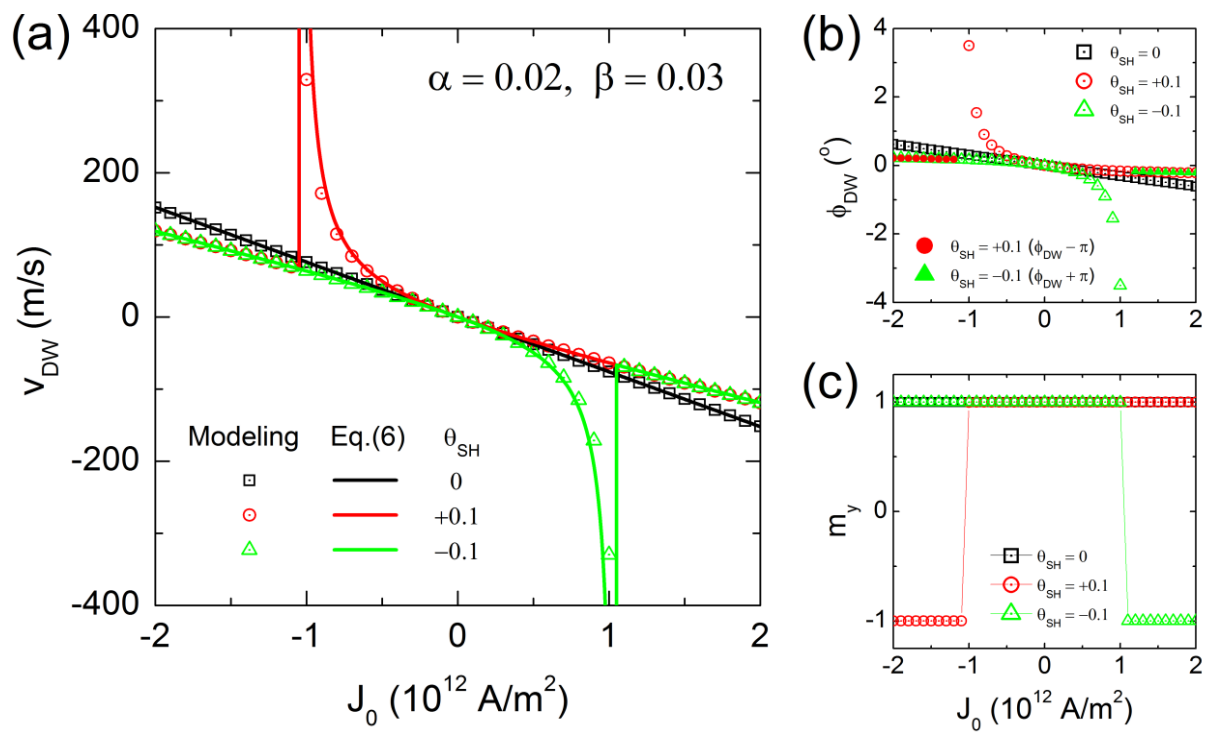

FIG. 3. Seo *et al.*